\documentclass[10pt]{llncs}
\usepackage{psfig}
\def\be{\begin{equation}}
\def\ee{\end{equation}}

\begin{document}

\title{Quantum Computation With Linear Optics}
\author{C. Adami and N. J. Cerf}
\institute{W. K. Kellogg Radiation Laboratory\\ California Institute 
           of Technology, Pasadena, California 91125, USA}

\maketitle

\begin{abstract}
We present a constructive method to translate small quantum circuits
into their optical analogues, using linear components of present-day
quantum optics technology only.  These optical circuits perform
precisely the computation that the quantum circuits are designed for,
and can thus be used to test the performance of quantum algorithms.
The method relies on the representation of several quantum bits by a
single photon, and on the implementation of universal quantum gates
using simple optical components (beam splitters, phase shifters,
etc.).  The optical implementation of Brassard et al.'s teleportation
circuit, a non-trivial 3-bit quantum computation, is presented as an
illustration.
\end{abstract}

\section{Introduction}
The promise of ultrafast computation using quantum mechanical
logic raised by Shor's discovery of a polynomial algorithm for
factoring~\cite{bib_shor} has yet to materialize in physical
implementations. While quantum logic has been implemented in a number
of different guises~\cite{bib_ex1,bib_ex2}, the dynamics and behavior
of a quantum circuit subject to noise and quantum decoherence has only
been tested in simulations on a classical
computer~\cite{bib_obenland,bib_paz} (but see 
\cite{bib_error}). There is little controversy about the
realization that it is the quantum mechanical superposition principle,
and the entangled, nonlocal, states it engenders, that are at the
origin of the speed-up of quantum algorithms with respect to their
classical counterparts. Still, effective quantum algorithms are few
and far between, and even those that are known today have yet to be
tested in a physical realization (but see~\cite{bib_alg}.)

In anticipation of physical realizations that implement quantum
superpositions between physical states, we present here a
method of constructing circuits based on non-local superpositions of
``eventualities'', rather than physical objects. More precisely, we
{\em simulate} quantum superpositions, ``qubits'', as ``which-path''
eventualities in linear optics, implementable on standard optical
benches. While the ``support'' of these qubits is decidedly classical
(the optical devices such as beam splitters, polarizers, etc.) the
wave function at the exit of the optical circuit can be made to coincide
arbitrarily well with the outcome of the anticipated computation,
thus implementing the quantum circuit. Naturally, this
``classical'' implementation of quantum logic has its drawbacks, as we
comment on further below. Still, it should provide an excellent (and
cost effective) means for testing small circuits for quantum error
correction or quantum algorithms.

As we point out below, the realization that optical which-path
eventualities of single photons can simulate qubits is not new in itself. 
Here, we focus on {\em protocols} to translate {\em any} quantum
circuit diagram into {\em linear} optics networks, which puts the
realization of simple circuits decidedly within reach. 
Quantum computation can be described as the task of performing 
a specific unitary transformation on a set of quantum bits (qubits) 
followed by measurement, so that the outcome of the
measurement provides the result of the computation.
This unitary transformation can be constructed with a finite number
of $4\times 4$ unitary matrices, that is, using a quantum circuit utilizing
only 1-bit and 2-bit quantum gates (see, {\it e.g.},
\cite{bib_qcircuit,bib_barenco}). The universality of 1- and 2-bit
gates in the realization of an arbitrary quantum computation was shown
in~\cite{bib_universal}.
Furthermore, it was realized recently
that an {\em  optical} realization
exists for any $N\times N$ unitary matrix~\cite{bib_reck}, a result which
generalizes the well-known implementation of $U(2)$ matrices
using a lossless beam splitter and a phase shifter
(see, {\it e.g.}, \cite{bib_U2}). Accordingly,
each element of $U(N)$ can be constructed using
an array of ${\cal O}(N^2)$ beam splitters that form an optical multiport
with $N$ input and $N$ output beams. As we shall see below, this
result together with the universality of (1- and 2-bit) gates, 
can be exploited {\em constructively}, providing a  
systematic method for assembling 
optically-simulated gates to build simple quantum circuits.
\par

\section{Logical Qubits in Optics}

Let us start by considering the equivalence between
traditional linear optics elements (such as beam splitters or
phase shifters) and 1-bit quantum gates (see, {\it e.g.}, \cite{bib_chuang}).
This equivalence is inspired by the standard  two-slit experiment of
quantum mechanics, in which a single quantum can interfere with itself
to produce fringes on a screen. Accordingly, a quantum on the other
side of the slit is in a superposition of paths, and the quantum
mechanical uncertainty principle is in full effect with respect to 
location and phase~\cite{bib_feyn}.

For example, in quantum circuit terminology,
an optical symmetric beam splitter is known to act
as a quantum $\sqrt{\rm \sc NOT}$ gate (up to a phase of $\pi/4$)
if we use the pair of input modes $|01\rangle$ (or $|10\rangle$) to represent
the logical 0 (or 1) state of the qubit. If one input port is in 
the vacuum state $|0\rangle$ (absence of a photon) and the second one
is in a single-photon
state $|1\rangle$, the output ports will then be in a superposition state 
$|01\rangle +i |10\rangle$. Thus, with the identification of the
{\em logical} qubits $|0_L\rangle \equiv |01\rangle$ and $|1_L\rangle
\equiv |10\rangle$, we produce the wavefunction
\be
|\Psi\rangle = \frac1{\sqrt2}\left(|0_L\rangle + i
  |1_L\rangle\right)\;,
\ee
from the initial state $|0_L\rangle$ just by running a photon through a
beam splitter. (The factor $i$ arises from the $\pi/2$ phase shift between
the transmitted and the reflected wave in a 
lossless symmetric beam splitter~\cite{bib_pi/2}.)

\begin{figure}[t]
\caption{Example of optical simulation of basic quantum logic
gates. (a) Hadamard gate on a ``location'' qubit, using a lossless
symmetric beam splitter.
(b) Controlled-{\sc NOT} gate using a polarization rotator.
The location and polarization
are the control and target qubit, respectively. (c) Same as (b) but
the control and target qubits are interchanged by the use of a polarizing
beam splitter.}
\vskip 1 cm
\centerline{\psfig{figure=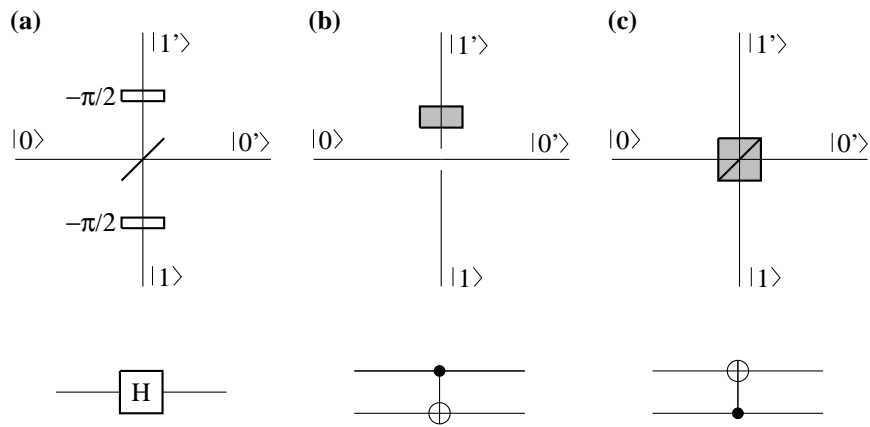,width=4.5in,angle=-90}}
\label{fig_intro}
\vskip 0 cm
\end{figure}

Similarly, a quantum phase gate
can be obtained by use of a phase shifter acting on one mode of the photon.
In other words, single-photon interferometry experiments 
can be interpreted in quantum circuit language,
the ``which-path'' variable being substituted with a quantum bit.
Although a general proof for the existence of an optical
realization of an arbitrary quantum circuit is implicitly given 
in Ref.~\cite{bib_reck}, the simple duality between 
quantum logic and single-photon optical experiments has not been exploited.
Here (and in \cite{bib_cak}) it is shown that a {\em single-photon}
representation of {\em several} qubits can be used to exploit this duality:
as several (say $n$) quantum bits can be represented 
by a single photon in an interferometric setup involving
essentially $2^n$ paths, quantum conditional dynamics can
easily be implemented by using different optical elements in distinct paths.
The appropriate cascading of beam splitters and other linear optical
devices entails the possibility of simulating 
networks of 1- and 2-bit quantum gates
(such as the Hadamard or the controlled-{\sc NOT} gate, 
see Fig.~\ref{fig_intro}), and thereby in principle achieving
universal $n$-bit quantum computations~\cite{bib_cak}.
This is in contrast with traditional optical mo\-dels
of quantum logic, where in general $n$ photons interacting
through {\em nonlinear} devices (acting as 2-bit quantum gates)
are required to represent $n$ qubits (see, {\it e.g.}, \cite{bib_chuang}).
Such models typically make use of the Kerr nonlinearity
to produce intensity-dependent phase shifts, so that the presence of
a photon in one path induces a phase shift to a second photon
(see, {\it e.g.}, the optical realization of a Fredkin 
gate~\cite{bib_milburn}). Instead, the method proposed here
yields a straightforward method for ``translating'' {\it any} $n$-bit
quantum circuit into a single-photon optical setup, 
whenever $n$ is not too large. The price to pay
is the exponential growth of the number of optical paths,
and, consequently, of optical devices that are required.
This will most likely limit the applicability of the proposed technique
to the implementation  of relatively simple circuits. 
\par

\begin{figure}[t]
\caption{Implementation of two sequential Hadamard transformations as
  a balanced Mach-Zehnder interferometer using lossless symmetric
  beam splitters only.}
\label{fig_mach}
\vskip 1 cm
\centerline{\psfig{figure=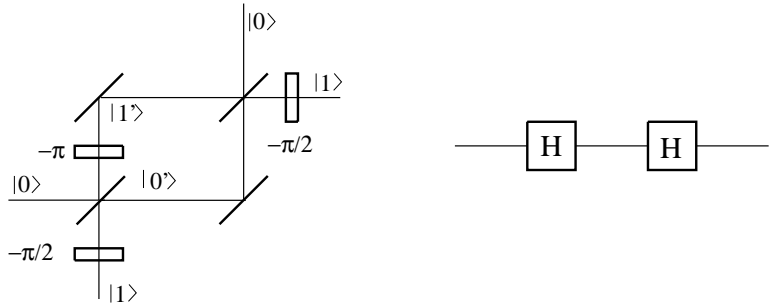,width=4in,angle=-90}}
\end{figure}

First, let us consider a single-photon experiment
with a Mach-Zehnder interferometer 
in order to illustrate the optical simulation of elementary quantum gates
(see Fig.~\ref{fig_intro}).
One qubit is involved in the description of the interferometer
in terms of a quantum circuit: the ``location'' qubit, characterizing
the information about ``which path'' is taken by the
photon. Rather than using the occupation number representation
for the photon, here we label the two input modes entering the
beam splitter by $|0\rangle$ and $|1\rangle$ (``mode description''
representation). The quantum state of the photon {\em exiting}
the beam splitter then is $|0'\rangle + i |1'\rangle$ or  
$|1'\rangle + i |0'\rangle$ depending on the input mode of the photon.
This is the $\sqrt{\rm \sc NOT}$ gate discussed earlier. Placing phase 
shifters at the input and output ports as shown in Fig.~\ref{fig_intro}a,
the beam splitter can be shown to perform
a Hadamard transformation between input and output modes, {\it i.e.},
\begin{equation}
\left( \begin{array}{c} |0'\rangle \\ |1'\rangle \end{array} \right)
= {1 \over \sqrt{2}}
\left( \begin{array} {cr} 1 & 1 \\ 1 & -1 \end{array} \right)
\left( \begin{array}{c} |0\rangle \\ |1\rangle \end{array} \right)  \;.
\end{equation}
In this sense, a lossless symmetric beam splitter (supplemented
with two $-\pi/2$ phase shifters) can be viewed as a Hadamard gate
acting on a location qubit.
Recombining the two beams using a second beam splitter (see
Fig.~\ref{fig_mach}) 
in order to form a balanced Mach-Zehnder interferometer corresponds 
therefore, in this quantum circuit language, 
to having a second Hadamard gate acting subsequently 
on the qubit\footnote{Here and below, 
it is understood that the path lengths are
adjusted so that the difference between dynamical phases vanishes.}.
Since $H^2=1$, it is not a surprise that the location qubit returns
to the initial basis state ($|0\rangle$ or $|1\rangle$) 
after two beam splitters (with the appropriate phase shifter). 
This sequence of two Hadamard gates 
simply conveys the fact that the contributions
of the two paths interfere destructively in one of the output ports, so that
the photon always leaves the interferometer in the other.
\par 

\begin{figure}[t]
\caption{Implementation of two sequential Hadamard transformations
  with intermediate {\em conditional} operation on the polarization.
 This circuit produces an entangled state $|0\rangle_{\rm pol}|0\rangle_{\rm
  loc}+|1\rangle_{\rm pol}|1\rangle_{\rm  loc}$
 between the polarization
 (which is initally in a product thate with the
 location qubit at the input port of the interferometer) 
 and the location qubit (denoted by $|0\rangle$ and $|1\rangle$ in the
 figure) just before the final beam splitter,
 preventing the observation of
 interference fringes at the output of this interferometer.}
\label{fig_mach2}
\vskip 1 cm
\centerline{\psfig{figure=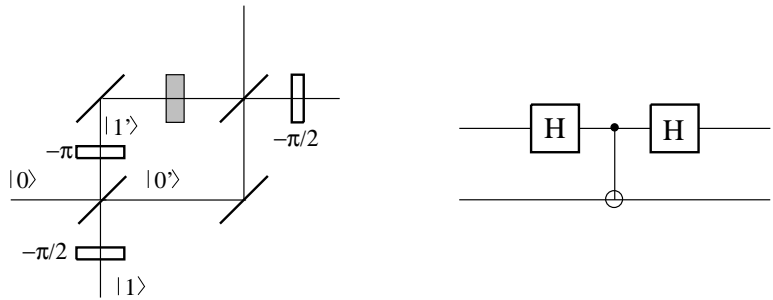,width=4in,angle=-90}}
\end{figure}

More interestingly,
consider now the same interferometer using polarized photons
(the photon is horizontally polarized at the input). Assuming that none
of the devices acts on polarization, the photon exits the interferometer
with the same polarization. In a circuit terminology, 
this corresponds to introducing
a ``polarization'' qubit ($|0\rangle_{\rm pol}$ 
stands for horizontal polarization)
which remains in a product state with the location qubit
throughout the circuit. If a polarization
rotator is placed in one of the branches of the interferometer, flipping
the polarization from horizontal $|0\rangle_{\rm pol}$ to vertical
$|1\rangle_{\rm pol}$,
it is well known that interference disappears since both paths
become distinguishable. This corresponds to placing
a 2-bit controlled-{\sc NOT} gate (represented in Fig.~\ref{fig_intro}b)
between the two Hadamard gates, where
the location qubit is the control and polarization is the target bit
(see Fig.~\ref{fig_mach2}).
The circuit in Fig.~\ref{fig_mach2} thus simply implements the
dynamics
\setcounter{equation}{0}
\renewcommand{\theequation}{3\alph{equation}}
\begin{eqnarray}
|0\rangle_{\rm pol}|0\rangle_{\rm
  loc}&\rightarrow& \frac1{\sqrt2}\left(|0\rangle_{\rm pol}|0\rangle_{\rm
  loc}+|1\rangle_{\rm pol}|1\rangle_{\rm  loc}\right)\\
&\rightarrow& \frac12\left(|0\rangle_{\rm pol}|0\rangle_{\rm
  loc}+ |1\rangle_{\rm pol}|0\rangle_{\rm
  loc}+|0\rangle_{\rm pol}|1\rangle_{\rm
  loc}-|1\rangle_{\rm pol}|1\rangle_{\rm  loc}\right)\,.\qquad  
\end{eqnarray}
which ``tags'' each path with a particular polarization just before
the final beam splitter in the sense that the polarization
of the photon is flipped {\em conditionally} on its location. The 
disappearance of interference fringes 
then simply reflects the entanglement between 
location and polarization qubits (the reduced density matrix  
obtained by tracing over polarization shows that the photon ends up in
a mixed ``location'' state, {\it i.e.}, it
has a 50\% chance of being detected in one or the other exit port). 
This suggests that Feynman's rule of thumb (namely that interference
and which-path information are complementary) is a manifestation
of the quantum {\em no-cloning} theorem: the location qubit cannot be
``cloned'' into a polarization qubit. However, the fringes can be resurrected
via a {\em quantum erasure} procedure~\cite{bib_eraser} (which involves placing
polarizing beam-splitters, introduced below,  at the exit ports of the
construction).

\par
The optical analogue of other basic quantum gates can be devised
following the same lines. For example, a polarizing beam splitter
achieves a controlled-{\sc NOT} gate where the location qubit is
flipped or not (the photon is reflected or not) conditionally on its
state of polarization, as shown in Fig.~\ref{fig_intro}c. 
Fredkin, Toffoli, as well as
controlled-phase gates can easily be simulated in the same manner but
will not be considered here. The central point is that, in principle, a
universal quantum computation can be simulated using these optical
substitutes for the universal quantum gates. The optical setup
is constructed straightforwardly by inspection of the quantum circuit.
A circuit involving $n$ qubits requires in
general $n$ successive splitting stages of the incoming beam,
that is, $2^n$ optical paths are obtained via $2^n -1$ beam splitters.
(The use of polarization of the photon as a qubit allows using
$2^{n-1}$ paths only.) This technique is thus limited to the simulation
of quantum networks involving a relatively small number of qubits
(say less than 5-6 with present technology).
The key idea of a quantum
computer, however, is to avoid just such an exponential size of the
apparatus by having $n$ physical qubits performing unitary transformations
in a $2^n$-dimensional space. In this respect,
it can be argued 
that an optical setup requiring $\sim 2^n$ optical 
elements to perform an $n$-bit quantum computation represents
a {\em classical} optical computer (see, {\it e.g.}, \cite{bib_barenco}).
Accordingly, the issue of whether non-locality (which is at the heart of 
entanglement) is {\it physically} present in the optical realization
is a matter of debate. 
\par

\section{Optical Quantum Teleportation}
\begin{figure}[t]
\caption{Quantum circuit for teleportation 
(from~\protect\cite{bib_teleport}). The initial state of qubit $\Lambda$ is
teleported to the state of qubit $\lambda$. Qubits $\sigma$ and $\lambda$
must be initially in state $|0\rangle$. Qubits $\Lambda$ and $\sigma$,
if measured at the end of the circuit, yield two classical (random)
bits that are uniformly distributed. }
\vskip 1 cm
\centerline{\psfig{figure=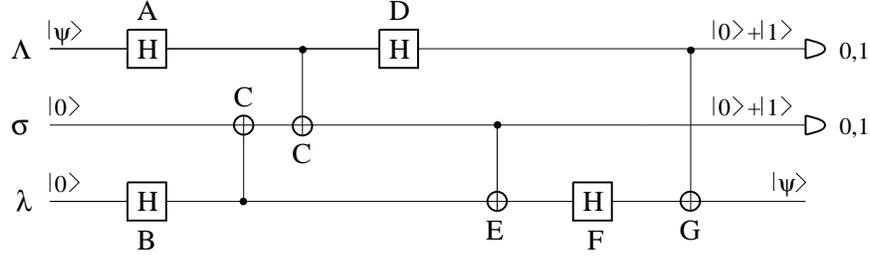,width=4.5in,angle=-90}}
\label{fig_circuit}
\end{figure}

As an illustration, we show that a quantum
circuit involving 3 qubits and 8 quantum gates
(see Fig.~\ref{fig_circuit}) 
can be implemented optically using essentially 9 beam splitters~\cite{bib_cak}.
This circuit\footnote{This teleportation circuit is equivalent to the one
described in \cite{bib_teleport}.}
has the property that the arbitrary
initial state $| \psi \rangle$ of qubit $\Lambda$ 
is ``teleported'' to the state in which qubit $\lambda$
is left after the process. In the original teleportation 
scheme~\cite{bib_originalteleport}, two classical bits (resulting from a Bell
measurement) are sent by the emitter, while the receiver performs
a specific unitary operation on $\lambda$ depending on these two bits.
However, it is shown in \cite{bib_braun} that these unitary operations
can be performed at the quantum level as well, by using quantum logic gates 
and postponing the measurement of the two bits to the end of the circuit.
The resulting quantum circuit (Fig.~\ref{fig_circuit})
is {\em formally} equivalent to the original teleportation scheme
(although no classical bits are communicated) as
exactly the same unitary transformations and quantum gates are involved.
While we do not claim that an optical realization
gives rise to ``genuine'' teleportation, this example circuit 
is instructive to demonstrate
the correspondence between quantum logic and optical devices
as it is small (3 qubits) but non-trivial.
\par

\begin{figure}[t]
\caption{Optical realization of the quantum circuit for teleportation
using polarized photons. The location qubit $\Lambda$ characterizes
the ``which-arm'' information at the first beam splitter, while
qubit $\lambda$ stands for the ``which-path'' information at the
second level of splitting. The initial location qubit $\Lambda$ is teleported
to qubit $\lambda$ and 
probed via the interference pattern observed at the upper
or lower ($\Lambda=0,1$) final beam splitter, for both polarization 
states ($\sigma=0,1$) of the detected photon. }
\vskip 0.5 cm
\centerline{\psfig{figure=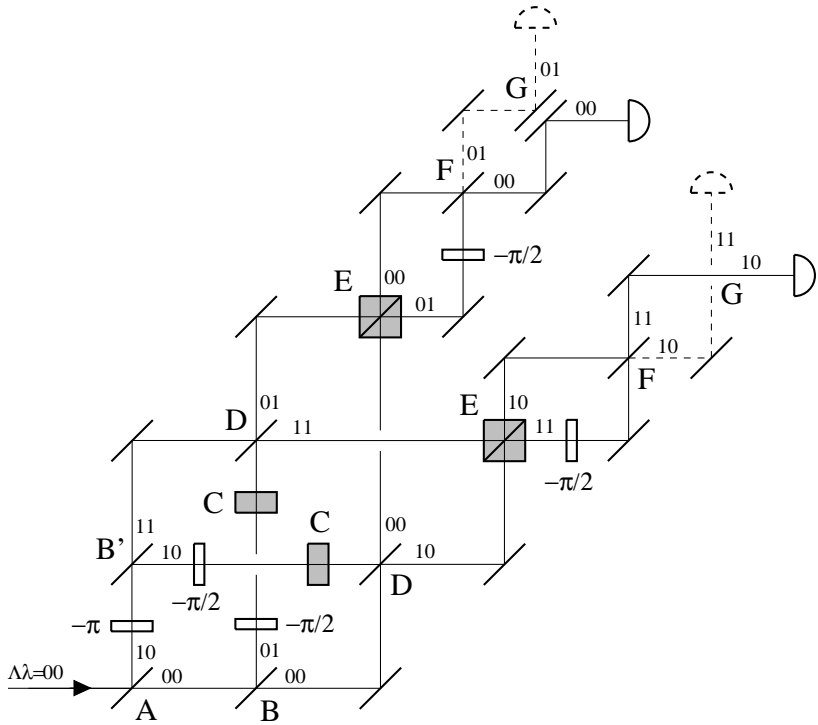,width=4.25in,angle=-90}}
\label{fig_optics}
\end{figure}

In the optical counterpart of this circuit (see Fig.~\ref{fig_optics}),
qubits $\Lambda$ and $\lambda$ correspond to the location of the photon
at the first and second splitting level, while $\sigma$
stands for the polarization qubit. Note that
the photons are initially horizontally
polarized, {\it i.e.}, in polarization state $|0\rangle$.
The first beam splitter A in Fig.~\ref{fig_optics} acts as a Hadamard gate
on $\Lambda$, as explained previously. 
For convenience, we depict the teleportation of state
$|\psi\rangle=|0\rangle$, so that
the incident photon enters this beam splitter in the input port
labeled $|0\rangle$. However, as any operation
in $U(2)$ can be realized optically, an {\em arbitrary} state of $\Lambda$
can be prepared (and then teleported) by having an additional
beam splitter (with tunable phase shifters) connected to
both input ports of beam splitter A.
The second level of beam splitters B 
(and B')\footnote{For convenience,
two realizations (B and B') of the Hadamard gate are used 
in Fig.~\ref{fig_optics}, where B' is obtained from B by interchanging
the $|0'\rangle$ and $|1'\rangle$ output ports
in Fig.~\ref{fig_intro}a.} corresponds
to the Hadamard gate B on $\lambda$ in Fig.~\ref{fig_circuit}.
The four paths at this point
($\Lambda \lambda =00$, 01, 10, and 11) label the four
components of the state vector characterizing qubits $\Lambda$ and $\lambda$. 
The probability amplitude for observing the photon in each of these
four paths, given the fact the photon enters the $|0\rangle$
port of beam splitters A and B, is then simply the corresponding
component of the wave vector. The combined
action of both controlled-{\sc NOT} gates C in Fig.~\ref{fig_circuit}
is to flip the
polarization state of the photon (qubit $\sigma$) conditionally on
the parity of $\Lambda+\lambda$ (mod 2), which is achieved by inserting
polarization rotators C at the appropriate positions. In other words,
the polarization is flipped on path 01 or 10, while it is unchanged
on path 00 or 11. 

The Hadamard gate D in Fig.~\ref{fig_circuit}
acts on qubit $\Lambda$, independently of $\lambda$. This is achieved
in Fig.~\ref{fig_optics}
by grouping the paths in pairs with the same value of $\lambda$
({\it i.e.}, crossing the paths) and using two beam
splitters D in order to effect a Hadamard transformation
on $\Lambda$ (one for each value of $\lambda$).
Similarly, the controlled-{\sc NOT} gate E acting on $\lambda$ (conditionally
on the polarization) is implemented by the use of two polarizing beam
splitters E after crossing the paths again. 
A polarizing beam splitter
leaves a horizontally polarized photon (state $|0\rangle)$) unchanged,
while vertical
polarization (state $|1\rangle$) is reflected.
The last Hadamard gate F in Fig.~\ref{fig_circuit}
corresponds to the last two
beam splitters F, and the final controlled-{\sc NOT} gate G
is simply achieved by crossing the paths ($\lambda=0,1$) in the lower arm
($\Lambda=1$) versus the upper arm ($\Lambda=0$).
In fact, the setup could be simplified by noting that
the conditional crossing of paths achieved by G simply reduces to
relabeling the output ports of beam splitter F in the $\Lambda=1$ arm.
In Fig.~\ref{fig_optics}, only those phase shifters associated
with the Hadamard gates (Fig.~\ref{fig_intro}a)
that are relevant in the final detection are indicated.

The interpretation of this optical circuit in the language of teleportation
is the following. After being ``processed'' in this quantum circuit,
a photon which was initially horizontally polarized can reach
one of the two ``light'' detectors (solid line in Fig.~\ref{fig_optics}) 
with horizontal or vertical polarization.
This corresponds to the final measurement of qubits $\Lambda$ and
$\sigma$ in Fig.~\ref{fig_circuit},
yielding two classical (random) bits: upper or lower arm,
horizontal or vertical polarization. The third qubit, $\lambda$, contains
the teleported quantum bit, that is, the initial arbitrary state of $\Lambda$.
Since the location state of the photon is initially $|0\rangle$
in the setup represented in Fig.~\ref{fig_optics},
it always exits to the ``light'' detector and never reaches the ``dark''
one (dashed line). For any measured value of $\Lambda$
(photon detected in the upper or lower arm) and $\sigma$ (horizontally
or vertically polarized photon), the entire setup forms a simple
balanced Mach-Zehnder interferometer. Indeed, there are exactly two
{\em indistinguishable} paths leading to each of the eight possible
outcomes (four detectors, two polarizations); these interfere pairwise,
just as in a standard Mach-Zehnder interferometer, explaining
the fact that the photon always reaches the ``light'' detector
(in both $\Lambda=0$ and $\Lambda=1$ arms and for both polarizations).
In this sense, the initial
``which-arm'' qubit $\Lambda$ has been teleported to the final
``which path'' qubit $\lambda$. 
Note that, as no
photodetection coincidence is required in this optical experiment,
the setup is actually not limited to {\em single}-photon interferometry.
This largely simplifies the realization of the optical source
since classical light fields (such as those from a laser) can be used rather
than number states.

\par
\section{Conclusion}
We have proposed a general technique for simulating
small-scale quantum networks using optical setups
composed of linear optical elements only. This avoids the recourse to
non-linear Kerr media to effect quantum conditional dynamics,
a severe constraint in the usual optical realization of quantum circuits.
A drawback of this technique
is clearly the exponential increase of the resources (optical devices)
with the size of the circuit. 
Nevertheless, as optical components that simulate
1- and 2-bit universal quantum gates can be cascaded straightforwardly,
a non-trivial quantum computing optical device can easily be constructed
if the number of component qubits is not too large. 
We believe this technique can be applied without fundamental difficulties
to the encoding and decoding circuits that are involved in the
simplest quantum error-correcting schemes~\cite{bib_qec}, 
opening up the possibility for an experimental simulation of
them. Furthermore, this technique promises a technologically simple 
way to test quantum algorithms for
performance and error stability. Last but not least, the
correspondence between quantum circuits and optical (interferometric)
setups suggests that new and improved interferometers could be
designed using the quantum circuit language~\cite{bib_jon}. 
\vskip 0.5cm 
\noindent{\bf \large Acknowledgments}
\vskip 0.25cm

\noindent We thank Paul Kwiat for many discussions and collaboration 
in this project.  This work was supported in part by NSF Grants PHY
94-12818 and PHY 94-20470, and by a grant from DARPA/ARO through the
QUIC Program (\#DAAH04-96-1-3086).  N.J.C.  is Collaborateur
Scientifique of the Belgian National Fund for Scientific Research.
\par

\end{document}